# Revealing intra-urban spatial structure through an exploratory analysis by combining road network abstraction model and taxi trajectory data

Sheng Hu[a, b, d], Song Gao[c, *], Wei Luo[c], Liang Wu[e, f, *], Tianqi Li[a], Yongyang Xu[b, e, f], Ziwei Zhang[a]

[a] *School of Geography and Information Engineering, China University of Geosciences, Wuhan 430074, China*
[b] *Guangdong-Hong Kong-Macau Joint Laboratory for Smart Cities, Shenzhen, 518061, China*
[c] *GeoDS Lab, Department of Geography, University of Wisconsin‐Madison, Madison, WI, USA*
[d] *Geography Department, National University of Singapore, Singapore, Singapore*
[e] *School of Computer Science, China University of Geosciences, Wuhan 430074, China*
[f] *National Engineering Research Center of Geographic Information System, Wuhan 430074, China*

*\* Corresponding authors:*
*Liang Wu, China University of Geosciences, Wuhan, P.R. China.*
E-mail: *wuliang@cug.edu.cn*
*Song Gao, Department of Geography, University of Wisconsin-Madison, 550 N. Park St., Madison, Wisconsin 53706-1404, USA.*
Email: *song.gao@wisc.edu*

**Abstract:** The unprecedented urbanization in China has dramatically changed the urban spatial structure of cities. With the proliferation of individual-level geospatial big data, previous studies have widely used the network abstraction model to reveal the underlying urban spatial structure. However, the construction of network abstraction models primarily focuses on the topology of the road network without considering individual travel flows along with the road networks. Individual travel flows reflect the urban dynamics, which can further help understand the underlying spatial structure. This study therefore aims to reveal the intra-urban spatial structure by integrating the road network abstraction model and individual travel flows. To achieve this goal, we 1) quantify the spatial interaction relatedness of road segments based on the Word2Vec model using large volumes of taxi trip data, then 2) characterize the road abstraction network model according to the identified spatial interaction relatedness, and 3) implement a community detection algorithm to reveal sub-regions of a city. Our results reveal three levels of hierarchical spatial structures in the Wuhan metropolitan area. This study provides a data-driven approach to the investigation of urban spatial structure via identifying traffic interaction patterns on the road network, offering insights to urban planning practice and transportation management.

## 1. Introduction

The unprecedented global urbanization in China has increasingly attracted the attention of urban planners and geographers to explore the intra-city structure with spatial interaction (e.g., daily commuting) (Barbosa et al. 2018; He et al. 2020; Liu et al. 2015; Martin & Schuurman 2020). Urban spatial structure is defined as the abstract of geographical space distribution (Chen et al. 2019; Foley 2016). A city acts as a dynamic and connected system with the spatial interaction flows of people between sub-regions (e.g., places and parcels) (Batty 2009). With different travel demands and people's behaviours, the resources and infrastructures of a city are allocated, resulting in the formation of complex and multi-dimensional city structures. Meanwhile, the rapid development of urban spatial structure has significantly impacted on urban dynamics and intra-city travel patterns of city's residents (Burger et al. 2014; van Meeteren et al. 2016; Wu et al. 2021). Over years, research efforts have been conducted to explore the impacts of city structure on the dynamic urban system (Yue et al. 2014). Revealing the underlying spatial structure can help suggest interesting insights into the organization and distribution of the urban system, which is important in terms of urban morphology, transportation, and traffic geography (Dokuz 2022; He et al. 2020; Liang & Kang 2021; Luo & MacEachren 2014; Schläpfer et al. 2021; Zhang et al. 2017).

With the proliferation of individual-level mobility data within cities, many complex network approaches have been implemented to reveal underlying spatial structures within a city (Batty 2013; Estrada, 2012; Lee & Kang 2015; Zhang et al. 2014;

Zhong et al. 2014). For example, the aggregated analysis models an entire city in a spatially embedded graph based on individual travel behaviours (De Montis et al. 2013; Liu et al. 2015; Yu 2018) However, the regional units (e.g., administrative districts and grid cells) in the aggregated analysis have been proved to be large and abstract, and detailed human movement information might be ignored without considering interaction relations between fine-grained spatial units, such as urban roads (Liu et al. 2017; Zhu et al. 2017). Recently, network abstraction models from the urban road network perspective are frequently applied (Hong & Yao 2019; Rodrigue et al. 2016). In the network abstraction model, the significant road intersections are represented as nodes and the road segments are represented as edges. The construction of the network abstraction model, however, primarily depends on the topology of the road network without considering the spatial interaction derived from individual travel flows. The integration of individual travel flows into the road network abstraction model can take advantage of the fine-grained spatial flow information among urban roads, bringing new insights to explore the underlying urban spatial structure.

Therefore, we proposed a novel framework to reveal the underlying spatial structure within a city by integrating the topology-based network abstraction model and individual travel flows derived from the taxi trajectory data. First of all, we used a representation embedding method-Word2Vec to quantitively measure the spatial interaction relatedness of road segments using taxi trajectory data. Then, we constructed a road network abstraction model based on the OSM road network, and further weighted the network using the quantified interaction relatedness. Finally, we employed the Infomap community detection method to reveal significant community patterns that represent sub-region patterns of a city. In addition, we used point-of-interest (POI) data to evaluate the spatial distribution of the detected regional structure of this city.

The remainder of this paper is organized as follows. Section 2 reviews the related literature on investigation of both urban spatial structure and road network abstraction model. Section 3 introduces the proposed framework and methods. In Section 4 and 5, we conducted a case study in the city of Wuhan and presented the results with discussions. Section 6 concludes this study and presents our vision on the future work.

## 2. Literature review

### 2.1. Revealing city structure based on the spatially embedded graph

The rapid growth of map services and mobile positioning technology has provided a massive amount of emerging geo-tagged data (e.g., taxi trajectory data and mobile phone record data) in the past decade. Those data sources lay the foundation to analyse the intra-urban flows to address many research issues, such as detecting human mobility patterns (Liu et al. 2012; Luo et al. 2018; Siła-Nowicka et al. 2016; Zhang et al. 2018; Zhu et al. 2018), assessing traffic indicators of the urban road network (Brauer et al. 2021; Cui et al. 2016; Hu, et al. 2021), revealing multi-level city structure (Gao et al. 2013; Liu et al. 2019; Liu et al. 2015; Zhong et al. 2014), and understanding urban land use (Liu et al. 2016; Liu, Wang, et al. 2012; Zhang et al. 2020)

Meanwhile, a spatially embedded graph (also called spatially embedded network) has widely been used to detect city structure with those human mobility data. Each geographical region stands for a node and the travel flow derived from the human mobility data stands for the edge between regions in a spatially embedded graph. Then, the community detection method is used to divide the entire graph into sub-graphs, namely communities or modules (Chen et al. 2015; Fortunato & Hric, 2016; Hric et al. 2014). Communities are sets of nodes that have strong inner connectivity and are sparsely connected to nodes of other communities. For example, Gao et al. (2013) empirically discovered the clustering structures of spatial-interaction communities according to the construction of two intertwined embedded networks, e.g., the network of phone-call interaction and the network of phone-users' movements. Liu et al. (2015) found a two-level polycentric community structure in Shanghai with grid cell based embedded graph using a taxi trip data. Therefore, significant community structures that represent the underlying sub-regions with strong internal travel flows can be revealed via spatially embedded graph (Hong & Yao 2019; Liu et al. 2015).

*2.2. Network abstraction model based on urban road network*

Recently, numerous in-depth researches have been conducted to reveal city structures via network abstraction model based on the urban road network. A road network, an artificial corridor in the urban areas, plays an important role in shaping the city's traffic and functional structure (Burghardt et al. 2021; Hong & Yao, 2019; Zhu et al. 2017). Inspired by the spatially embedded graph, the network abstraction model regards the urban road network as an embedded graph, in which the significant road nodes (e.g., intersections) are represented as nodes and the road segments are represented as edges. Hong & Yao (2019) empirically found a hierarchical structure of communities according to the network abstraction model based on OSM road network.

Spatially embedded graph models an entire city to explore the underlying city structure using individual travel flows, but the regional units are large and abstract. Network abstraction model primarily focus on the inner topology of the road network, but do not consider the role of human movements along with the road network. Previous studies tend to employ individual travel behaviours and road network abstraction model respectively. This research, therefore, aims to reveal the intra-urban spatial structure by combining road network abstraction model and individual travel flows.

*2.2. A critical analysis*

In sum, many efforts have been made for revealing the urban spatial structure, but they have the following shortcomings:

- The units to explore the underlying urban spatial structure using individual travel flows on spatially embedded graph models are aggregated to areal units that may suffer from the modified area unit problem (MUAP).
- Network abstraction model primarily focus on the inner topology of the road network, but do not consider the role of human movements along with the road network. Previous studies tend to employ individual travel behaviours and road network abstraction model separately.

This research, therefore, aims to reveal the intra-urban spatial structure by combining road network abstraction model and individual travel flows. We proposed an integrated framework to reveal the underlying spatial structure within a city based on individual travel flows derived from the taxi trajectory data. The contributions of this work are as follows:

- We proposed an integrated framework for sensing the underlying hierarchical urban spatial structure, which can benefit from both network abstraction model and detailed human movement information at a fine-grained scale.
- We investigated the integration of the quantified spatial interaction relatedness of road segments into the urban road network to enrich urban road network-based GIS research.

## 3. Methodology

In this study, we propose a novel framework to explore hierarchical urban spatial structure using urban road network and taxi trip data (Fig. 1). We start with measuring the spatial interaction relatedness of road links based on the Word2Vec model. A community detection method based on fine-scale traffic interaction among urban roads is then used to discover communities. The last step validates the detected results.

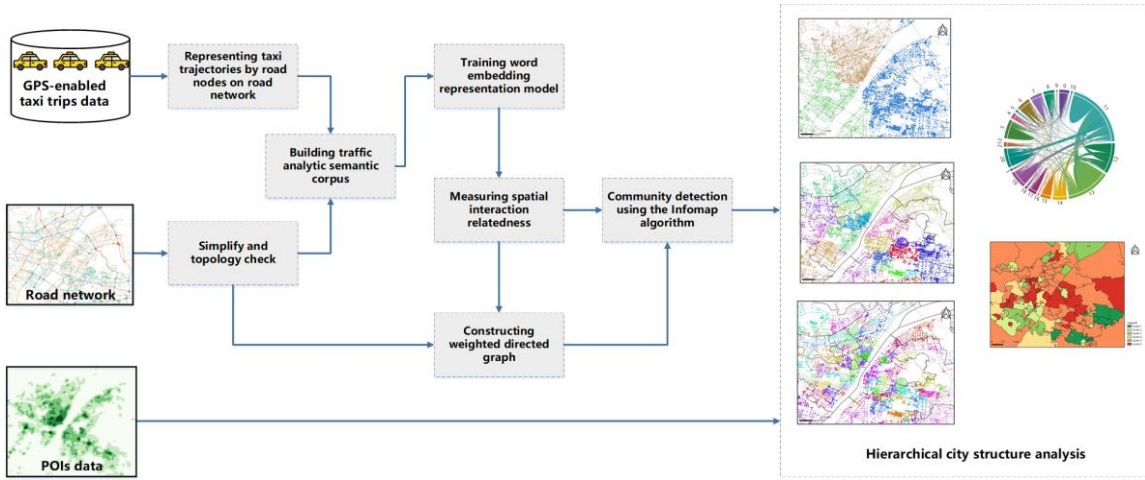

Fig. 1. The flowchart of the proposed framework.

## 3.1. *Measurement of spatial interaction relatedness of road segments based on the Word2Vec model*

In this section we introduce the Word2Vec model to quantify the spatial interaction relatedness of road segments on the urban road network using GPS-enabled taxi trajectory data.

### 3.1.1. *Representing taxi trajectories by road nodes on road network*

Taxis operate along the urban road network and taxi trip routes contain valuable human activity information about people's movement and traffic flow. Here, we present a map-matching-based approach to represent taxi trajectories along a road network using the interaction node as a spatial assembly and analytic unit. In this study, a road node defined as the point is used to represent connectivity between two road segments, including the starting point, the ending point, and the intersection of the road segments. Specifically, we start with mapping taxi GPS records to urban roads via a fast map matching algorithm (Yang & Gidófalvi 2018), and then we represent each route as the sequence of consecutive matched road nodes (Fig. 2).

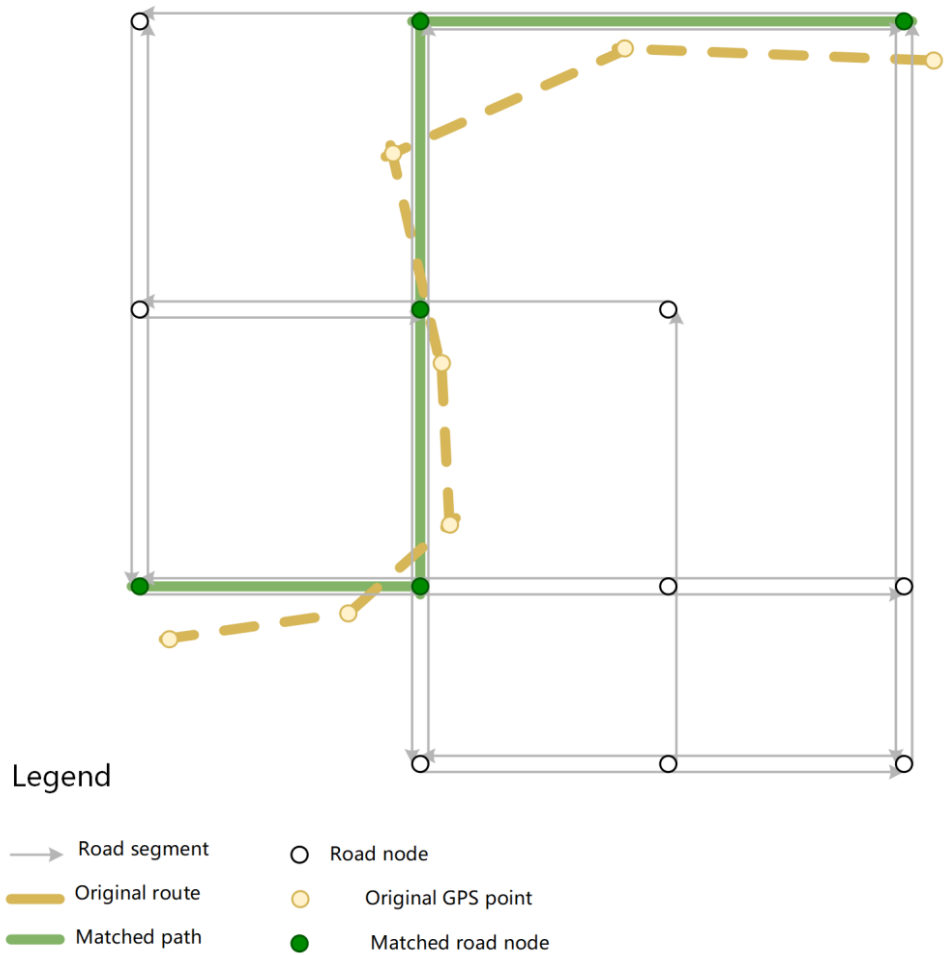

Fig. 2. An example of mapping a raw taxi route to intersection nodes sequence along the road network.

### 3.1.2. Building traffic analytic semantic corpus

In recent years, geo-semantic analysis framework derived from the natural language processing (NLP) field has shown potential as a promising tool to exploit spatial relationships in urban areas based on big data (Cai 2021; Luo et al. 2019; Yao et al. 2017; Liu, Gao, and Lu 2019; Crivellari and Ristea 2021). By analogizing traffic elements to NLP terms, geo-semantic analysis builds high-dimensional embedding vectors to quantitatively describe the traffic components, hence exploring potential information in geographical data (Bengio et al. 2013; Hu, Gao, et al. 2021). Researchers can effectively train a language model and evaluate potential semantical representations or semantical relationships based on a large semantic corpus in geo-semantic analysis framework.

In this study, based on the assumption that the traffic interaction indicates the travel activities of urban people and to be intimately connected with the urban spatial structure, the entire study area was analogized as a traffic analytic semantic corpus. In this corpus, the taxi movement trajectories are analogized to NLP documents, while road nodes on the road network are words in documents. The objective of building the corpus is to capture fine-scale traffic interaction patterns along with the urban road network and measure the spatial co-occurrence relationships between traffic nodes. A higher degree of traffic relatedness between two intersection nodes indicates that they are likely to co-occur in the taxi movement trajectories along with the road network.

*3.1.3. Training word embedding representation model*

Word embedding is one of the most popular language representation techniques of document vocabulary originated from the NLP domain. For machine learning tasks, the words must be represented meaningfully. As a result, they must be numerically stated. As word embedding techniques are used to solve such problems, algorithms such as Word2Vec enable words to be expressed mathematically (Cai 2021; Liang & Kang 2021). The core principle behind word embedding techniques is that each word is represented by a D-dimensional real-valued vector. This vector comes to reflect the semantics of a word and the distance between two vectors can be used to quantify the semantic similarity and semantic relatedness (Mai et al. 2022). The word embedding technique represents each distinct word with a numerical vector representation via a self-supervised neural network model. It is capable of capturing the context of vocabulary words in a document and further measuring the semantic relation (similarity and relatedness) with other words.

Word2Vec is an open-source, state-of-the-art tool proposed by Mikolov et al. (Mikolov et al. 2013) to achieve word embedding with ease of operation and high scalability. Word2vec trains a two-layer neural network and reconstructs linguistic contexts of words for converting each word to a unique vector. In this study, the skip-gram based Wrod2vec model (Levy & Goldberg 2014) was introduced to obtain the numerical representation of intersections (as nodes) along with the road network. Given a consequence of road nodes (a trajectory route), the goal of the skip-gram based Word2vec model is to predict context road nodes for a given target node in a sliding window. The maximum likelihood function can be estimated as:

$$J(\theta) = \frac{1}{n}\sum_{j=1}^{n} \log p\left(v_j | v_{j-w}^{j+w}\right) \quad (1)$$

where $w$ is the training context window size, $n$ is the number of road nodes, and $v_{j-w}^{j+w}$ denotes the context road nodes of the target one $v_j$. The conditional probability $p(v_j|v_{j-w}^{j+w})$ can be estimated as:

$$p(v_j|v_{j-w}^{j+w}) = \frac{\exp(v_j, v_{j-w}^{j+w})}{\frac{1}{n}\sum_{j=1}^{N} \exp\left(v_j, v_{j-w}^{j+w}\right)} \quad (2)$$

We train the Word2vec model using the traffic analytic semantic corpus, reconstruct spatial contexts of traffic elements, and finally obtain the symbolized real-valued vector representation for each road node.

*3.1.4. Measuring spatial interaction relatedness*

The urban road network constrains the majority of human movement, and vehicle routes contain valuable human activity information about people movement and traffic flow hidden in the urban road network. The measurement of relatedness (Liu et al. 2017), cosine similarity, between two road nodes aims to quantitively investigate the inherent spatial interactions on a road network, and consequently revealing intra-urban spatial structure. If two nodes in the road network have a strong traffic relatedness, they are more likely to co-occur along taxi trajectories. A cosine similarity measurement is employed to measure the traffic relatedness of the road nodes based on their embedding vectors, denoted as (Xu et al. 2021):

$$sim(w_i, w_j) = \frac{w_i \cdot w_j}{\| w_i \| \| w_j \|} \quad (3)$$

where $sim(w_i, w_j)$ means cosine similarity between embedding vectors $w_i$ and $w_j$ and changes between 0 to 1. A value of $sim(w_i, w_j)$ approaching 1 indicates that road node $i$ has strong correlations with road node $j$. It is worth noting that the obtained similarity denotes the topological measurement of urban roads and quantitates the functional relations of road interactions (Hong & Yao 2019).

### 3.2. Identification of underlying spatial structure using the Infomap algorithm

#### 3.2.1. Road network abstraction model

Road network architectures in urban settings have been studied through the analogy of urban road networks to graphs in complex network disciplines (Gao et al. 2013; Liu et al. 2015; Yao et al. 2021). With the emergence of OSM, researchers are able to better explore urban road networks and access urban spatial interaction patterns. In this study, the road network abstraction model was constructed using OSM road data. Specifically, the road network was symbolized to a weighted directed graph $G \equiv (V, E, W)$, where the road node is referred to as graph vertex, $V$, the road linking two adjacent nodes is referred to as an edge, $E$, the road driving direction is used to describe the edge direction, and the spatial interaction relatedness of the road segment is used to describe the weight of edge, $W$. The construction of the road network abstraction model was implemented using an open-sourced tool-OSMnx (Boeing 2017).

*3.2.2. Community detection using the Infomap algorithm*

The community detection method attempts to group or divide graph vertexes into a few subsets based on their interaction patterns. The divided sets of vertexes, commonly called communities or modules, are strongly linked to each other and sparsely connected to the rest of the graph (Xie et al. 2013). Based on the spatial interaction relatedness and topological connectivity of road nodes, the Infomap community detection algorithm is employed to reveal significant community patterns that represent sub-region patterns of a city. Among community detection algorithms, the Infomap method with its superior performance and high availability has emerged as a popular method in a wide range of applications (Liu et al. 2015; Yao et al. 2021).

InfoMap algorithm is a network partition approach for modularity analysis built on the minimum entropy principle and random walks strategy. Network partition, given a graph, refers to a specific division of the nodes into modules with an objective function. Infomap optimises the objective function known as the map equation (Edler et al. 2017). Infomap uses Huffman coding (Huffman 1952) to describe each node in the network with a two-level description and minimizes the description length of a random walker's movements over all possible network partitions on a network. That is, the partition with the shortest description length best describes the community structure of the network in terms of network dynamics. More details about Infomap can be found in Rosvall & Bergstrom (2008). In this study, the Infomap algorithm is used to partition the urban road network into multiple layers in order to show the network's rich hierarchical characteristics. The input of the Infomap algorithm is the weighted directed graph $G$, and the output is a divided graph $G$ in which each node is grouped into hierarchical modules/subgroups.

*3.2.3. Mixed land used indices*

Inspired by Yue et al. (2017), three mixed-use indices, including the Margalef Species Richness index, Shannon's entropy index, and the Simpson's index of diversity, are introduced to measure the extent of mix/evenness in the distributions of land use types. The Margalef Species Richness index (Gotelli & Colwell 2011), namely Richness index, is a species diversity index in the ecology domain and is used to estimate the degree of diversity in the land use and POI context. The Richness index for the specific region $i$ is defined as $R_i$:

$$R_i = \frac{S-1}{\ln n_i} \quad (4)$$

where $S$ is the number of POI types and $n_i$ is the number of POI within the region $i$. Note that a lower value of $R_i$ indicates the less diversity of POI types, suggesting a higher purity urban function of the division.

The Shannon's entropy index (Brown et al. 2009), namely Shannon index, reflects the degree of orderliness of the distribution of POI types, which is defined as $H_i$:

$$H_i = -\sum_{j=1}^{S} p_j \ln p_j \quad (5)$$

where $p_j$ is the percentage of the number of POI type $j$ in total number of POIs of region $i$. Note that A lower value of $H_i$ indicates the greater orderliness and less random of the distribution of POIs.

The Simpson's index of diversity (Hunter & Gaston 1988; Simpson 1949) measures the

concentration of the distribution of POIs, which is defined as $D_i$:

$$D_i = 1 - \frac{\sum_{j=1}^{S} m_j(m_j - 1)}{n_i(n_i - 1)} \qquad (6)$$

where $m_j$ is the total number of POIs of a particular type $j$ within the region $i$; $n_i$ is the total number of POIs of all types within region $i$. Note that a higher value of $D_i$ indicates a greater degree of concentration of region $i$. The above three indices evaluate mixed land use in the POIs context from different aspects, and therefore can effectively verify the advantage of our proposed method in delineating urban functional areas.

## 4. Case study

### *4.1. Study area and data collection*

A case study was carried out in the city of Wuhan, China that is known for its complex urban morphology and high rates of mixed land use. Because of Wuhan's rapid urbanization, new requirements for exploring the urban spatial structure and urban planning have been emerging in recent years. The study area was selected from a downtown area of Wuhan city, covering an area of 748.39 km² (Fig. 3). The study area is characterized by a wide variety of growth and population density conditions, in which the landscapes are exceedingly diverse.

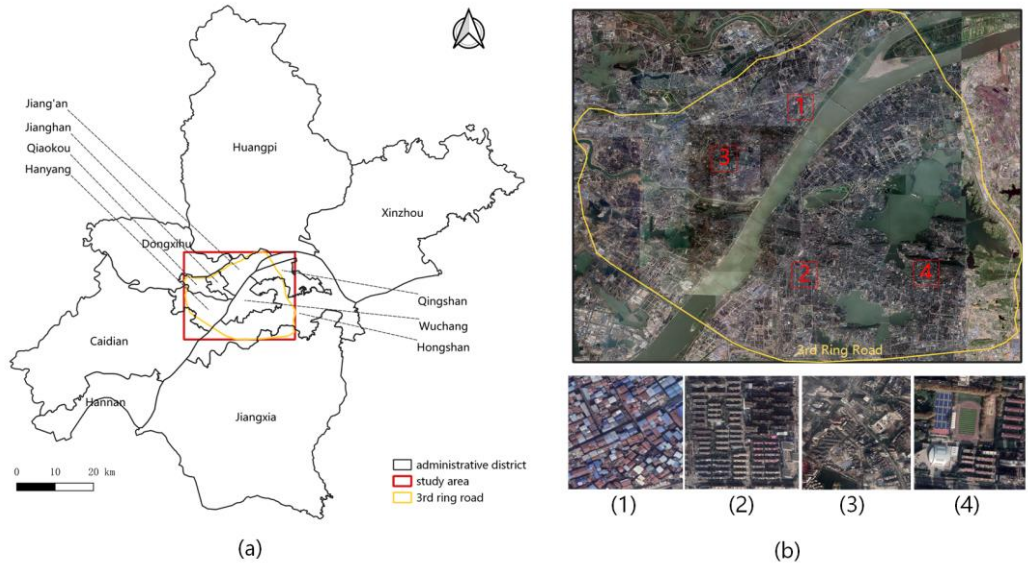

Fig. 3. Our study area in Wuhan, which covers the downtown area of this city (within the third ring road). (a) Administrative districts of Wuhan city; (b) Satellite remote sensing image of the study area. The bottom images show examples of diverse landscapes: (1) industrial area, (2) residential area, (3) commercial area and (4) educational area.

In this work, three kinds of geographic data, including urban road network, taxi GPS trajectories, and POI data, were utilized to conduct the experiments. Note that the coordinate system of all geospatial data is unified to the WGS geographic coordinate system.

- **Urban road network**. The primary road network data was obtained from OSM, which was acquired in January 2018. OSM is an open-source map service that gives users free and easy-to-access digital map data, and it is now the most popular and successful volunteered geographic information provider (Xu et al. 2019). The road data contains essential attribute information, including road name, road type, coordinate location (longitude and latitude) of road nodes, and topological connection information. Extraordinarily high precision in the position and topological relationship is found in the study area. 22,115 road segments and

14,715 nodes were extracted after pre-processing procedures such as simplification and topological checking (Fig. 4).

- **GPS-enabled taxi trajectories**. As an important form of transportation, GPS-enabled taxis are not constrained by routes or time and provide a kind of flexible and wide-ranging trajectory data in urban regions with high accuracy and fewer privacy concerns (Zheng et al. 2011). Taxi GPS record data used in this study was collected from May 9, 2015-May 15, 2015 in Wuhan, which contains movement tracks of more than 10,000 taxis. The original taxi trajectory dataset is essentially a collection of GPS track points. Each point includes basic driving data such as taxi number, timestamp, coordinate location, speed, direction, and status (vacant or occupied). The sampling frequency of the GPS track is about 50 s. With the attribute of occupied status, passenger-carrying trajectories are retrieved from the original GPS track points. A consecutive taxi movement route is composed of the pick-up location, the drop-off location, and several intermediate GPS records.

- **POIs**: The POI data was collected from an open-source data platform, Peking University Research Data (State Information Center 2017). This data covers the period from July 1, 2017 to September 30, 2018. A total of 404,721 POIs is extracted in the study area (Fig. 5). A POI consists of essential attributes: POI identity number, name, hierarchical types, address, and coordinate location. Hierarchical types are grouped into three categories, i.e., top-level, second-level, and third-level. From the top level to third level, the POIs descriptions are provided in greater detail. In this study, to balance the trade-off between the mixture of urban land use and the complexity of semantic computation, the second-level category of POIs types is utilized to identify urban functions and evaluate the effectivity. POI data is mainly used to calculate the land use mixed indices and identify the urban functions in hierarchical city structure analysis.

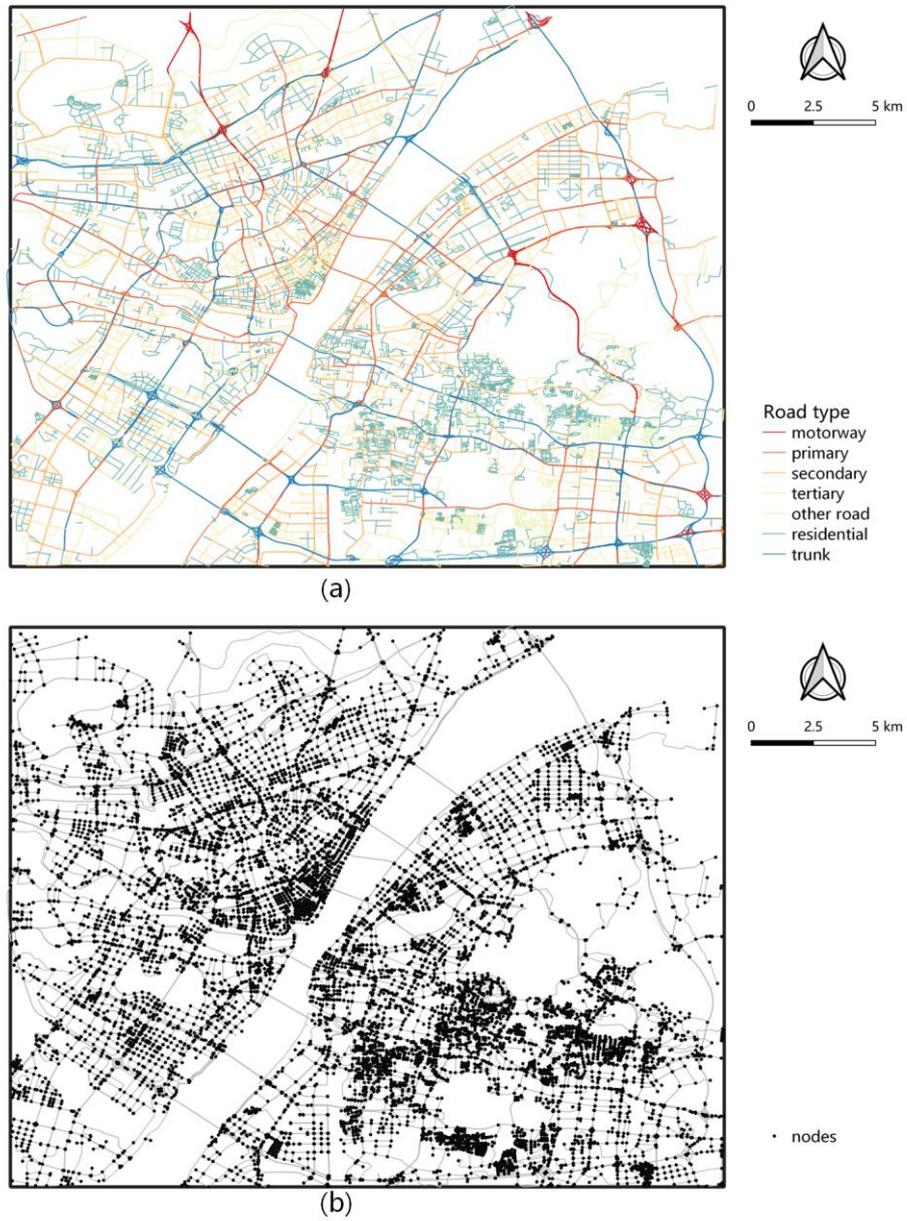

Fig. 4. Data schema of the road network in the study area. (a) road segments and (b) road nodes.

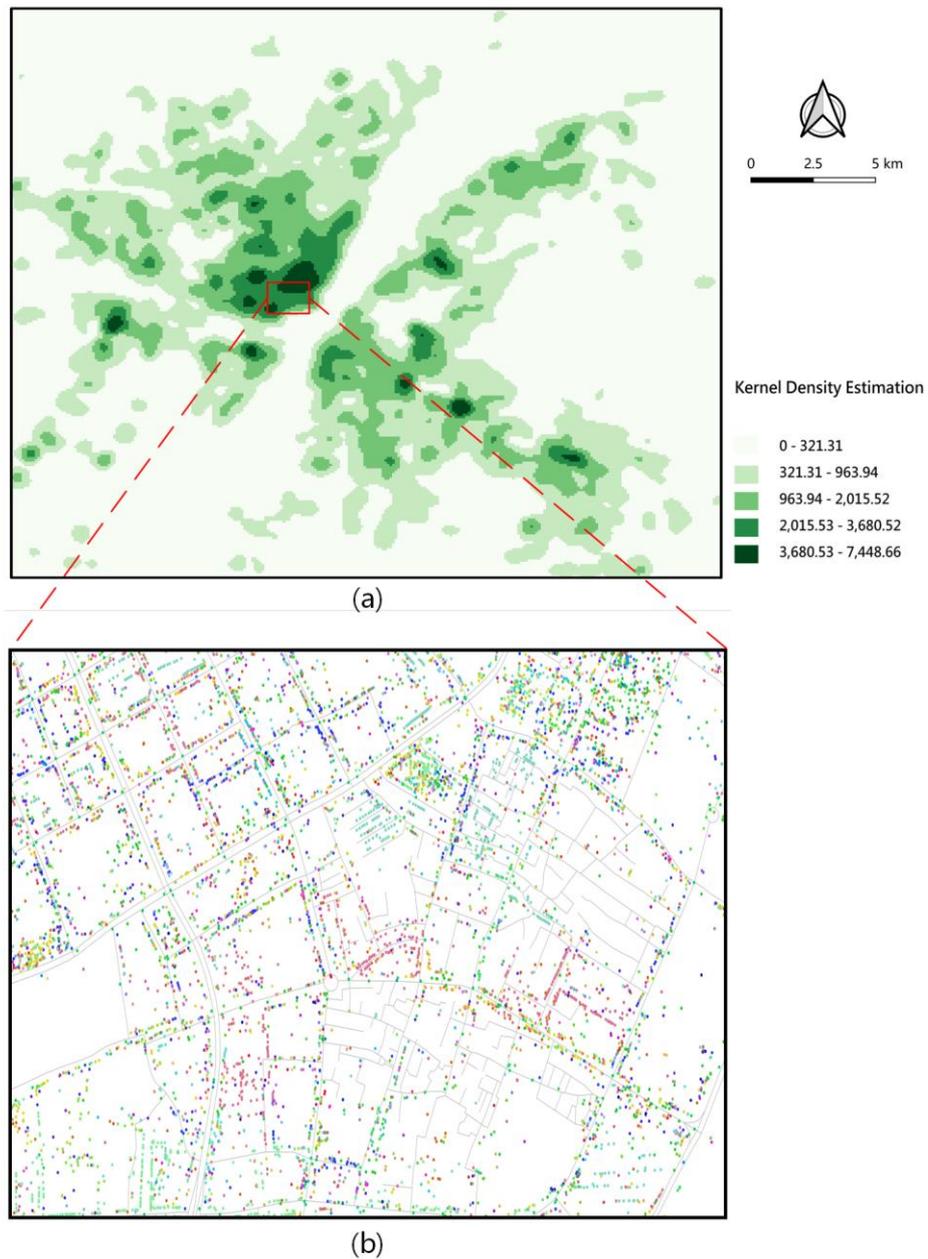

Fig. 5. (a) kernel density estimation of POIs (unit: count per km$^2$); (b) spatial distribution of POIs near Jianghan Road, a famous commercial area within the downtown of Wuhan. Note that different colours indicate types of POIs.

### *4.2. Results*

#### *4.2.1. Embeddings representation and spatial interaction relatedness of road nodes*

The road network in the study area was abstracted onto a weighted directed graph. Taking the road network graph as input, a skip-gram based Word2vec model was trained to obtain embedding representations. In view of the volume of graph structure and the computational costs, most parameters were set to recommended or default values. Specifically, for model training, the dimension of the road node representation was set to 128, the number of training epochs was set to 10, and the window size was set to 10. The Word2vec model was trained using the representation learning tool-gensim in Python (Rehurek & Sojka 2010).

Cosine similarity metrics were then utilized to obtain the relatedness between any two road nodes of a road segment using the embedding vectors. Fig. 6 demonstrates the relationship between the average similarity and the distance between pairwise two node embeddings. For each road node, we randomly selected 10% of all road nodes to calculate the average similarity and road distance between two nodes. We found that the topological correlation generally follows Tobler's first law of geography (Tobler 1970): near nodes are more related than distant nodes. Meanwhile, Fig. 7 shows an example of the spatial distribution of the similarity of road nodes in a local area. The red node denotes selected centre node. As the blue colour deepens, so does the similarity between the centre node and its neighbours, implying a larger topological correlation. The nodes on roads with the same direction are more related than roads in the opposite direction, which indicates that the similarity of road nodes can be influenced by traffic interactions.

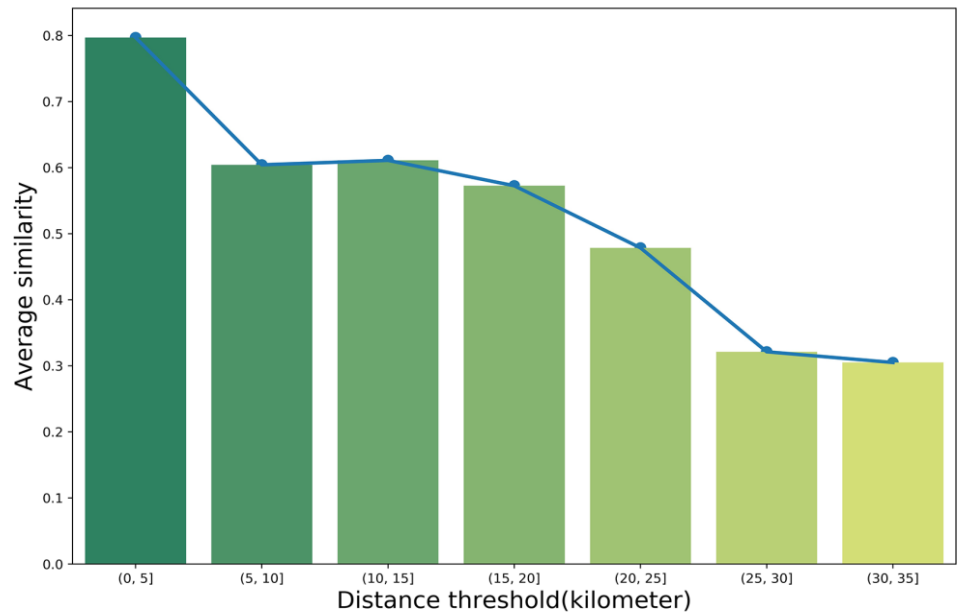

Fig. 6. Relationship between the average similarity of node embeddings and the distance between two nodes.

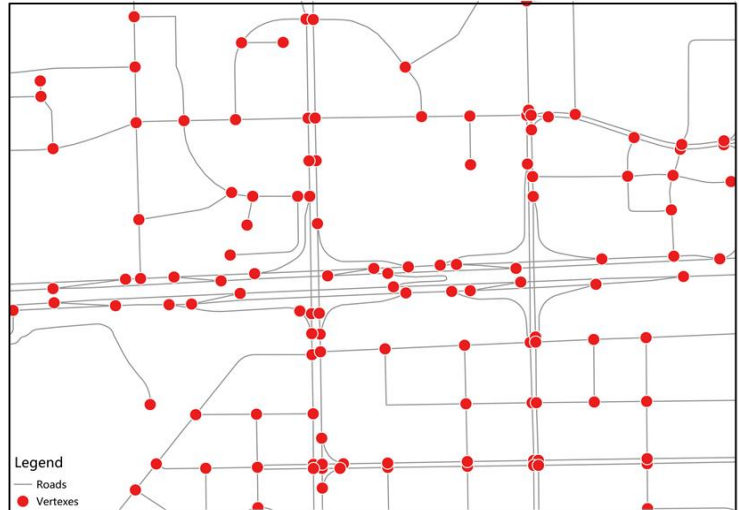

(a)

(b)

(c)

Fig. 7. An example of embedding vectors in a road network graph: (a) road networks; (b) embedding vectors of road nodes; (c) spatial distribution of cosine similarities.

*4.2.2. Hierarchical urban spatial structure*

Based on spatial interaction relatedness and topological connectivity, the Infomap algorithm was further employed to implement multilayered detection of communities, and hierarchical urban spatial structure was further explored. 错误!未找到引用源。 shows the statistical results of three-level communities. At the top level, the entire urban road network in the study area is divided into three aggregated sub-regions. As depicted in Fig. 8a, the division result is very consistent with the previous urban spatial pattern, namely *Three Towns of Wuhan*, which formed in the last century before the administrative regionalization reform in China ("Wuhan" 2021). *Three towns of Wuhan* consist of *Wuchang*, *Hankou*, and *Hanyang*, which are three independent towns and located in three disjoint parts in the downtown area of Wuhan. Each town has its own development characteristics and patterns with respective specialties in economy, culture and industry. This result indicates that the spatial structure of the *Three Towns of Wuhan* still has far-reaching effects on the contemporary urban spatial structure.

Table 1. Comparison of different communities.

| Different communities | Top-level | Second-level | Third-level |
| --- | --- | --- | --- |
| Total number of communities | 3 | 22 | 127 |
| Average edge weight (relatedness) | 0.857 | 0.859 | 0.864 |

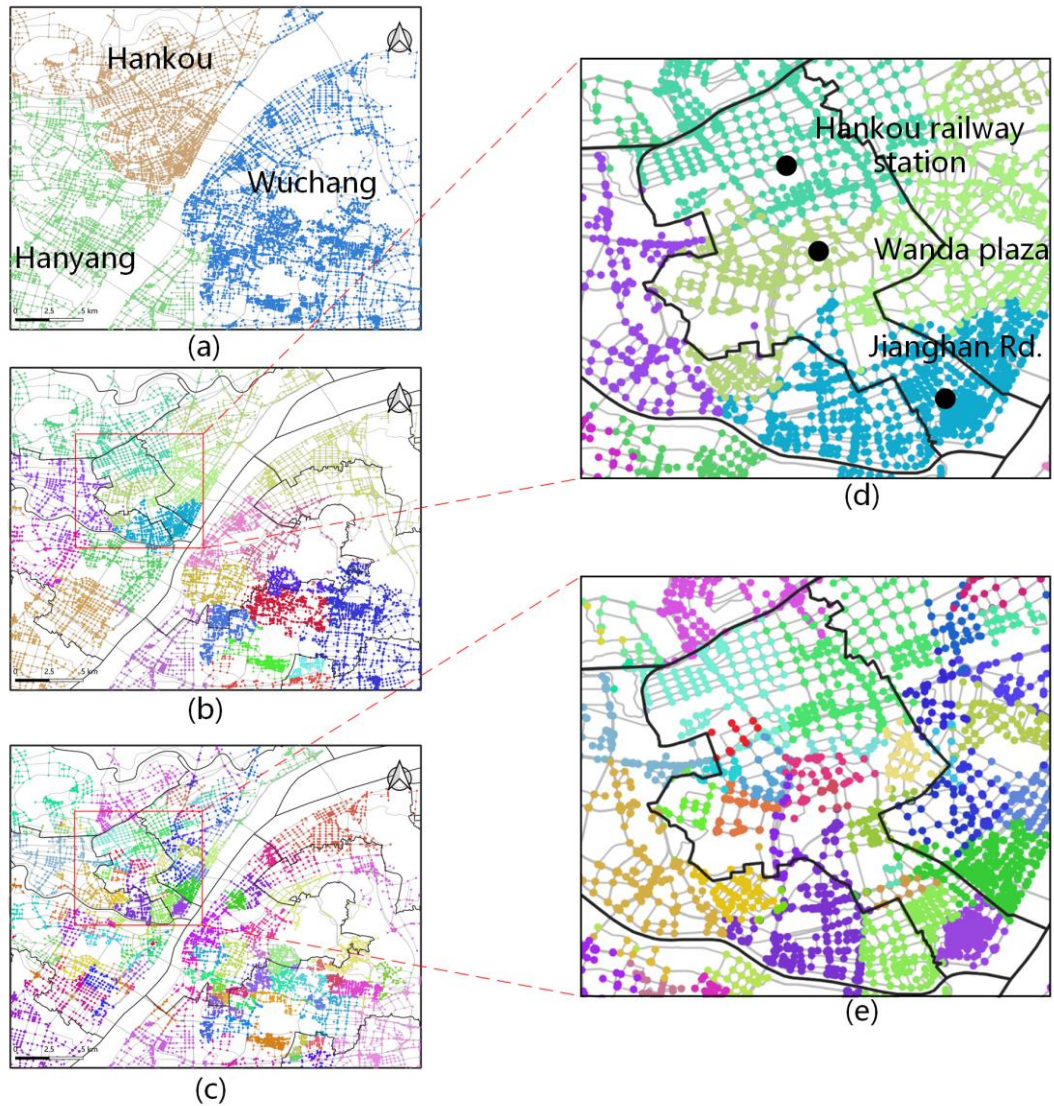

Fig. 8. The spatial distribution of hierarchical communities. Note that the solid black lines indicate the administrative boundaries. (a) top-level communities; (b) second-level communities; (c) third-level communities; (d) and (e) indicate the second-level divisions and third-level divisions of Jianghan district.

As shown in Fig. 8b, more detailed structural information is detected at the second-level communities. The study area is divided into 22 communities or sub-regions with an average edge weight of 0.859. Most studies usually demonstrate the sub-regional structure of a city by referring to its administrative divisions (De Montis, Caschili, and Chessa 2013; Liu et al. 2015; Ratti et al. 2010). In the study area, most sub-regions are inconsistent with the district-level administrative division, indicating the different connectivity between the districts and the underlying urban spatial structure via intra-city interactions in the study area. For example, Jianghan district, located in the downtown area of Wuhan, is mainly divided into three different regional areas, corresponding to three different functional areas (Hankou railway station transportation area, Wanda plaza commercial area, and Jianghan Road shopping area) (as shown in Fig. 8d). These three areas with different functions have distinct spatial interactions among each other. Administrative division structure is a static artifact which is partially arbitrary. Cities function as dynamic systems where traffic flows play a significant role in connecting the discrete resources of a city into an integrated system. Our results embrace the same idea that spatial interaction ties sub-regions within a city.

To further validate our results and understand the spatial interaction patterns between the second-level communities, the chord diagram was used to visualize the actual traffic flows using the origins and destinations of taxi trips. As depicted in Fig. 9, we found that the frequency of intra-flows is greater than inter-flows between sub-regions, indicating that most taxi movements happen within the same sub-regions. This result shows that the proposed methods can effectively reveal regional urban spatial structure and identify the intra-city spatial interaction patterns.

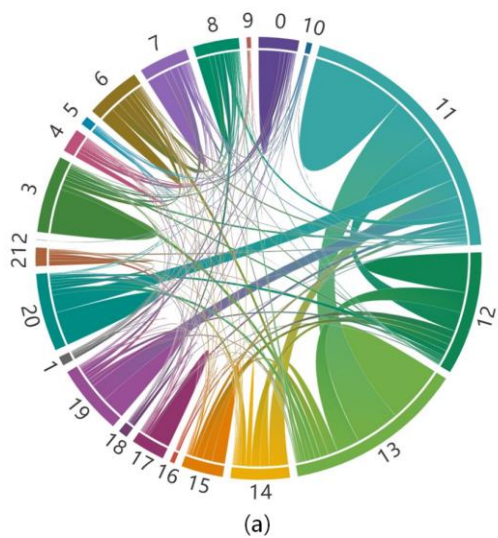
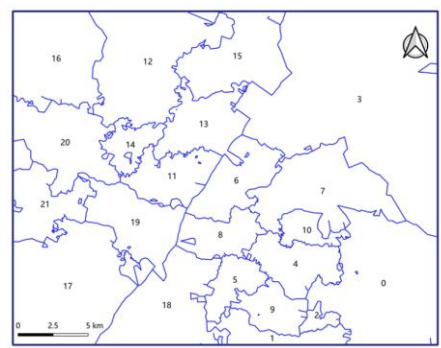

(a) (b)

Fig. 9. The chord diagram of traffic flows using origins and destinations of taxi movements. Noting that the numbers in the left diagram and right map is consistent and indicate the identifier of the sub-regions. (a) the chord diagram; (b) the spatial distribution of sub-regions. The blue lines indicate the sub-regions boundary generated by Thiessen polygon algorithm.

Fig. 8c maps a more elaborate regional division pattern with 127 communities and an average edge weight of 0.864. These results reveal a fine-scale urban spatial structure, and each sub-region can reflect specific urban functional areas. We employed the Thiessen polygon to estimate the influence area of each road node and determine the coverage area of each division. To further explore the functional area characteristics, the term frequency-inverse document frequency (TF-IDF) method was used to identify the urban functions using POI types. By weighting different POI types within each sub-region, the TF-IDF method stresses the uniqueness of POI types in each sub-region and decreases the weight of common POI types, therefore effectively identifying functional characteristics of each sub-region. A similar approach has been used in previous works (Gao, Janowicz, and Couclelis 2017; Hu et al. 2021; Liu et al. 2020); detailed descriptions about TF-IDF algorithm can be found in Beel et al. (2016).

Fig. 10 shows the spatial distribution of urban functional areas, and 错误!未找到引用源。 shows the function descriptions. Cluster 1 mainly includes educational areas such as many universities. Cluster 2 consists of many kinds of comprehensive marketplaces. Cluster 3 includes most industrial areas located alone with 3rd Ring Road. Cluster 4 includes business and commercial areas located near Wuhan Central Business Unit (CBD). Cluster 5 is heavily mixed areas of residential blocks and commercial areas. Transportation, few business and industrial areas dominate Cluster 6.

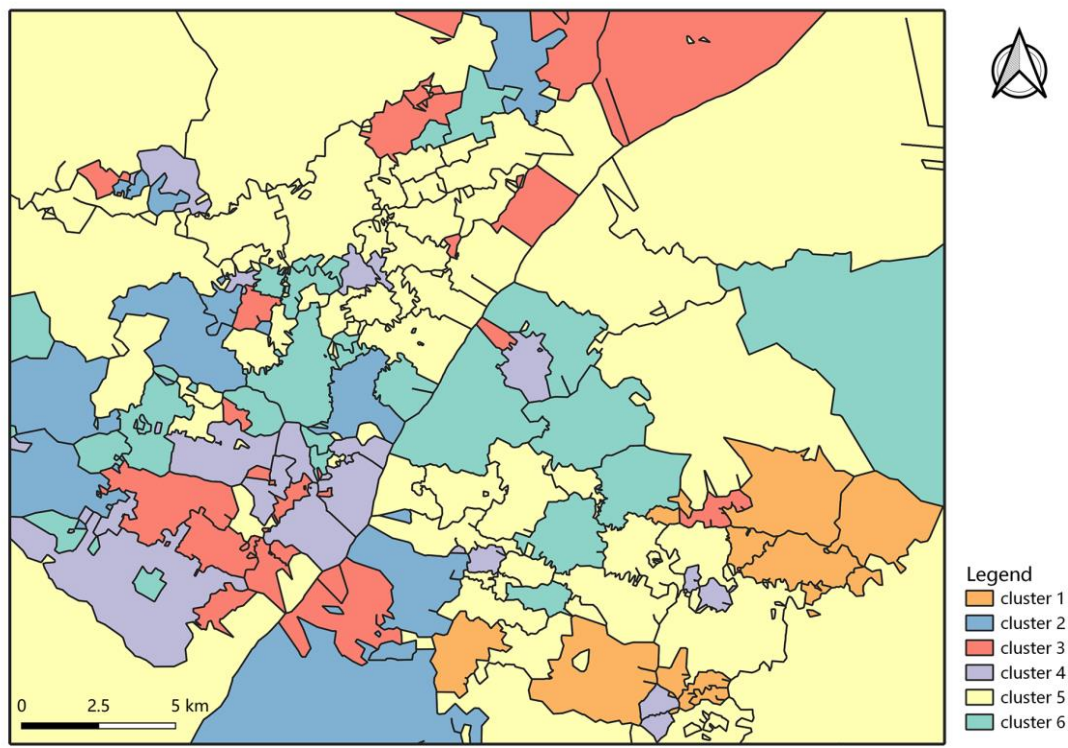

Fig. 10. The spatial distribution of urban functional areas based on third-level division.

Table 2. Urban functions areas and example locations.

| Sub-region cluster | Function types | Example locations |
| --- | --- | --- |
| Cluster 1 | Educational area | Huazhong University of Science and Technology |
| Cluster 2 | Comprehensive market area | Qingling Road Materials Market, Wuhan Furniture Store |
| Cluster 3 | Industrial area | Hongqiao Industrial Park |
| Cluster 4 | Business, commercial area | Wuhan International Expo Center |
| Cluster 5 | Residential, commercial area | Donghu Ecological Tourism and Scenic Area, Gangdu Residential Area |

| Cluster 6 | Transportation, business and industrial area | Wuchang Railway Station, Acer Bus Transport Station |

*4.2.3. Result verification*

To evaluate the effectiveness of our proposed method, we implement two comparison experiments with different weights of each road segment within the urban road graph:

- **O-Infomap**: Original Infomap method with all weight of edges set to 1;
- **D-Infomap**: Distance-weighted Infomap method with weight set as Euclidean length of each road segment;
- **Our proposed method**: Relatedness-weighted Infomap method with weight set as spatial interaction relatedness between traffic nodes.

Three hierarchical spatial division patterns were detected using the comparison experiments in a quantitative way. As suggested by Brown & Holmes (1971) and Noronha & Goodchild (1992), one classic approach to measure the performance of delineating urban functional regions is maximizing the interactions within the same region while minimizing the interactions between different regions. Inspired by Liu et al. (2019), we calculated the percentage of average frequency (PAF) of taxi flows between the second-level divisions within the same division over seven days. A greater value of PAF suggests that people are more inclined to travel within the same division or sub-region, indicating a more significant spatial division and better capability of identifying spatial interaction patterns. As shown in 错误!未找到引用源。 the values of PAF are higher than that of the other methods on most days, proving that our proposed method performs significantly better than other methods in identifying spatial interaction patterns.

Table 3. The percentage of average frequency of taxi flows between the second-level divisions within the same division of seven days.

| Methods | May 9, 2015 | May 10, 2015 | May 11, 2015 | May 12, 2015 | May 13, 2015 | May 14, 2015 | May 15, 2015 |
|---|---|---|---|---|---|---|---|
| O-Infomap | 0.294 | 0.288 | 0.312 | 0.292 | 0.299 | 0.297 | 0.300 |
| D-Infomap | 0.339 | 0.369 | 0.351 | 0.374 | 0.359 | 0.382 | 0.366 |
| Proposed method | 0.379 | 0.372 | 0.391 | 0.378 | 0.378 | 0.384 | 0.386 |

Moreover, we calculated multiply indices for mixed land use of the third-level divisions using POI types. We hypothesize that the mixing degree of urban functions in a specific sub-region reflects regional urban functional structure, thus verifying the validity of the division results. A lower mixing degree of urban functions suggests a more pronounced spatial division and is better capable of delineating urban functional areas.

Table shows statistical information of three mixed-use indices using different methods. Our proposed method achieves good fitting results with a lower average Richness index and average Shannon index, indicating the advantage of our proposed method in keeping the purity of urban functions and orderliness of the distribution of land-use types. However, compared to other methods, Simpson's index of diversity has no significant difference. In summary, the quantitative comparison results from two different perspectives indicate the effectiveness of our proposed approach in revealing hierarchical spatial structure in a city.

Table 4. Indices for mixed land use of the third-level divisions.

| Methods | $R_i$ | $H_i$ | $D_i$ |
|---|---|---|---|

|  | Number of divisions | Mean | Std. | Mean | Std. | Mean | Std. |
| --- | --- | --- | --- | --- | --- | --- | --- |
| O-Infomap | 124 | 105.236 | 368.350 | 8.859 | 35.313 | 0.945 | 0.087 |
| D-Infomap | 147 | 55.508 | 102.843 | 4.014 | 6.694 | 0.953 | 0.036 |
| Proposed method | 127 | 37.417 | 64.454 | 3.156 | 5.614 | 0.956 | 0.037 |

**5. Discussion**

*5.1. Hierarchical urban spatial structure*

Urban spatial structure is closely related to the urban travel patterns generated from their daily lives of urban dwellers. The distribution of various urban elements (or infrastructure) significantly impacts people's travel, and their movement flows connect discrete parcels or areas into an integrated system. Benefited from the spatial interaction analysis, urban parcels or areas, such as grids, communities, or traffic analysis zones could be joint via the similarity of functionality to obtain spatial interactions among aggregated regions (i.e., such urban functional region). With multilayer aggregating at different scales, hierarchical urban spatial structure could be revealed (Hong & Yao 2019; Wu et al. 2021). For example, by dividing taxi trips into short- and long-distance trips, Liu et al. (2015) revealed a two-level hierarchical urban spatial structure based on the different travel demand of local dwellers and their mobility patterns. In this hierarchical urban spatial structure, local clusters are generated with short-distance trips (dominate local spatial interactions), whereas long-distance trips play the role of connecting these local clusters. Our results resonate with Liu's typical work, and particularly address the spatial hierarchy at a fine-grained scale.

Exploring the hierarchical urban spatial structure have great meanings in multiple application domains. Hierarchical urban spatial structure highlights the multilayer spatial interactions, therefore could bring new insights for extracting the co-location patterns and underlying spatial semantics of urban functional regions. What's more, different hierarchical levels of urban regions associated with their functions could be dynamically drawn on a digital map (i.e., tourist map or urban planning map). Such hierarchical semantical map could be beneficial for various map users (Gao 2017). For city planners, it is critical to answer questions such as 'where are the most dynamic regions in the city' and ' where are the neighborhoods that are changing the most '? Moreover, hierarchical urban spatial structure could reveal the regions that attract a lot of attention from the general public in an urban area. As a result, when limited resources are available for urban planning initiatives, these regions may be given a higher priority.

### *5.2. The spatial interaction relatedness and human mobility patterns within a city*

There has been continued and sustained interest in extracting hidden information from human mobility patterns in urban studies. Previous studies generally extract temporal or spatial statistical characteristics (i.e., frequencies statistics of origins and destinations of the individual-level trips) to determine an urban region within a city. Origins and destinations represent the travel purposes of people, but the intermediate trajectory information between origins and destinations is rarely used (Hu et al. 2021; Zheng et al. 2014; Zhu et al. 2017). Inspired by Hu et al. (2021), an analogy strategy from urban elements to NLP field was introduced, where the urban area was regarded as a corpus, a travel trip was deemed as a sentence, road node was used as word. Further a Word2Vec model was employed to measure the spatial interaction relatedness between road nodes.

Embracing the analogy strategy from GIS elements to the NLP field, considerable research in recent years has demonstrated the advantages of word embedding technologies in GIScience from different perspectives. For example, Crivellari (2021) firstly introduced the CrimeVec approach based on word embedding technology to understand the criminology of urban place. Word2Vec model was used to creating dense vectors of crime types based on spatial-temporal distribution. Zhang et al. (2020) proposed a Traj2Vec model to quantify trip trajectories as high-dimensional semantic vectors using mobile phone positioning data. In addition, they proved that cell towers with similar vectors are spatially closer to one another. Attempts have also been made to measure traffic interactions (Liu et al. 2017; Wu et al. 2020; Xu et al. 2020), delineate urban functional use (Huang et al. 2022; Niu & Silva, 2021; Sun et al. 2021) and so on using word embedding models (Mai et al. 2022).

The spatial interaction relatedness has the unique advantage in extracting the underlying human mobility patterns. Compared with traditional origins and destinations information, the spatial interaction relatedness thoroughly extracts refined characteristics by integrating fixed urban road network structure and dynamic human movement trips. In this regard, the spatial interaction relatedness could be used to reveal the movement flows and traffic states. For example, how upstream or downstream road traffic flows impact the neighborhood roads? We could employ the embedding representation and cosine-based similarity measurement to quantify the impact.

## 6. Conclusions

It is one of the central themes for transport geographers to reveal the underlying urban spatial structure from the intra-urban spatial interaction patterns. In this study, we employed the Word2Vec model to quantify fine-scale interaction relations of roads segments on a road network using taxi trajectory data. We then utilized the Infomap community detection method to reveal significant regional patterns in the Wuhan metropolitan area. We found the three-level hierarchical structures that indicate the division of urban space with the integration of a large volume of real taxi trajectories into the urban road network. Urban space is divided into three aggregated sub-regions at the top level, which shows great consistency with the previous urban spatial pattern called Three Towns of Wuhan. The second level reflects sub-regions with more intra-flows than inter-flows. The third level reveals a fine-scale spatial distribution of urban functions. We further assessed the advantage of our proposed method via two comparison experiments. Our work considers the entire city as a connected system using human travel flows on the urban road network. Both our method and results can provide traffic managers and urban planners with a better understanding of traffic patterns on the road network and the regional structure of a city.

Furthermore, we would like to regard this research as a beginning of detecting the spatial interaction communities based on trajectory data and road network abstraction model. Further research can be conducted to enhance our proposed framework and compare with state-of-the-art methods if more detailed urban geographical data and open sources are available. Future work is also anticipated to collect multimodal trip data (e.g., subway, biking, bus trips) to support a more comprehensive spatial interaction investigation in cities.

**Disclosure statement:**

No potential conflict of interest was reported by the authors.

**Data Availability Statement:**

The data and codes that support the findings of this study are available from the corresponding author upon reasonable request.

**Acknowledgments:**

This work was supported by the National Natural Science Foundation of China [grant number 41871311]; Guangdong Science and Technology Strategic Innovation Fund (the Guangdong-Hong Kong-Macau Joint Laboratory Program, Project No. 2020B1212030009); and the Fundamental Research Funds for National Universities, China University of Geosciences (Wuhan); China Scholarship Council (CSC) during a visit of Sheng Hu to National University of Singapore.